\documentclass{article}

\usepackage[english]{babel}

\usepackage[letterpaper,top=2cm,bottom=2cm,left=3cm,right=3cm,marginparwidth=1.75cm]{geometry}

\usepackage{amsmath}
\usepackage{graphicx}
\usepackage[colorlinks=true, allcolors=blue]{hyperref}
\usepackage{authblk}

\title{Transmission of optical communication signals through ring core fiber using perfect vortex beams}
\author[1]{Nelson Villalba}
\author[2]{Crist\'obal Melo}
\author[3,4]{Sebasti\'an Ayala}
\author[2]{Christopher Mancilla}
\author[2]{Wladimir Valenzuela}
\author[2]{Miguel Figueroa}
\author[1]{Erik Baradit}
\author[5]{Riu Lin}
\author[6]{Ming Tang} 
\author[3,4]{Stephen P. Walborn}
\author[3,4]{Gustavo Lima}
\author[2]{Gabriel Saavedra}
\author[1,*]{Gustavo Cañas}

\affil[1]{\small{Departamento de F\'isica, Universidad del B\'io-B\'io, Collao 1202, 5-C Concepci\'on, Chile}}
\affil[2]{Departamento de Ingenier\'ia El\'ectrica , Universidad de Concepci\'on, 160-C Concepci\'on, Chile}
\affil[3]{Departamento de F\'{\i}sica, Universidad de Concepci\'on, 160-C Concepci\'on, Chile}
\affil[4]{Millennium Institute for Research in Optics, Universidad de Concepci\'on, 160-C Concepci\'on, Chile}
\affil[5]{Department of Electrical Engineering, Chalmers University of Technology, 41296 Göteborg, Sweden}
\affil[6]{Wuhan National Lab for Optoelectronics (WNLO) \& National Engineering Laboratory for Next Generation Internet Access System, School of Optical and Electronic Information, Huazhong University of   Science and Technology, Wuhan 430074, China}
\affil[*]{gcanas@ubiobio.cl}

\begin{document}
\maketitle

\begin{abstract}
Orbital angular momentum can be used to implement high capacity data transmission systems that can be applied for classical and quantum communications. Here we experimentally study the generation and transmission properties of the so-called perfect vortex beams and the Laguerre-Gaussian beams in ring-core optical fibers. Our results show that when using a single preparation stage, the perfect vortex beams present less ring-radius variation that allows coupling of higher optical power into a ring core fiber. These results lead to lower power requirements to establish fiber-based communications links using orbital angular momentum and set the stage for future implementations of high-dimensional quantum communication over space division multiplexing fibers.
\end{abstract}

\section{Introduction}

To increase the capacity of optical fiber communication systems, the physical properties of light such as wavelength (frequency), polarization, amplitude, and phase have been used to multiplex information through the years \cite{Zhu:18, Kan:17, Winzer:17, Gnauck:11, Galdino:20, Ionescu:19, Hamaoka:18, Renaudier:17}. Nevertheless, due to the nonlinear nature of silica-based optical fibers, data capacity is reaching its current limits \cite{Essiambre:08, Essiambre:10}. To date, transmission of $244.3$ Tb/s over $54$ km has been achieved using all the aforementioned physical dimensions \cite{Puttnam:22}. To further increase the information carrying capacity of optical fiber, new degrees of freedom need to be exploited, where the use of an extended fiber bandwidth and additional spatial modes are some of the most promising solutions \cite{Winzer:17, Puttnam:22}. Space division multiplexing (SDM) has been introduced as a viable way to multiplex channels in the spatial-domain, and novel fiber types have been developed to support transmission of multiple spatial modes. Current examples of fibers  that support SDM include multi-core fibers, few- and multi-mode fibers, and ring-core fiber (RCF) \cite{Richardson:13, Sillard:14, Rademacher:18, Saitoh:16}.

The orbital angular Momentum (OAM) of light can be associated to optical beams with helical phase-fronts described by $e^{il\theta}$, where $\theta$ is the azimuthal angle in cylindrical coordinates and $l$ can take integer values and it is usually called topological charge\cite{Allen:92}. The intensity profile from OAM modes is characterised by a doughnut shape, since  destructive interference along its propagation axis generates a dark zone.  OAM multiplexing has been a subject of intense research lately, with great interest in free-space and fiber-based optical communications systems \cite{Yao:11, Willner:21, Minghao:21, Pang:21}. OAM  eigenmodes with different $l$ values form an orthogonal basis in the associated Hilbert space. Moreover, the number of OAM modes can be infinite in principle, so that they present the potential to increase the information capacity carried per photon \cite{Erhard:18,willner:17}.

To date, OAM has been used to establish classical communications links and transmission of $1.6$~Tb/s was achieved over a $1.1$~km fiber link using OAM multiplexing \cite{Bozinovic:13} and $100$~Tb/s in free-space data link enabled by three-dimensional multiplexing of orbital angular momentum, polarization, and wavelength \cite{willner:17}. 
The main difficulties in implementing OAM-based SDM systems is propagation and coupling of OAM modes into an optical fiber. 
First, to propagate OAM modes in an optical fiber, the fiber should present a large effective index separation between vector modes and a refractive index profile compatible with the OAM field \cite{Brunet:15}. Both of these conditions can be achieved using ring-core fiber (RCF) \cite{Brunet:15, Jung:17, Jin:16}. Second, coupling a large number of OAM modes into a RCF remains a challenging task. For free-space transmission, the Laguerre-Gaussian (LG) modes are a natural choice, as they maintain their transverse structure during free-space propagation.  However, the doughnut-shaped intensity profile of these modes depends on the topological charge of the beam. For this reason, most of the experimental demonstrations of OAM propagation in RCFs use two modes either with the same topological charge with values $\pm l$  \cite{Bozinovic:13}, or different topological charge with similar beam diameter \cite{Long:18}. This limitation determines the practical implementations of large OAM-based SDM systems, and thus the capacity increases attainable by this technique.
To solve this problem, the use of perfect vortex beams (PVB) has been proposed theoretically to efficiently couple OAM modes in RCF fiber \cite{Rojas:21}. PVBs are OAM modes that present a nearly constant intensity radius regardless of the topological charge \cite{Ostrovsky:13,Garcia:14}, and can be mathematically obtained by Fourier transformation of Bessel-Gauss modes \cite{Vaity:15}.

In this article, we experimentally study the coupling and transmission of OAM modes using RCFs and evaluate their transmission properties for telecommunication and quantum applications. We used both PVB and LG modes in order to evaluate performances in classical and quantum communications, benchmarked by the bit error rate in the classical case, and by the average quantum state fidelity in the quantum case.

\section{OAM generation and fiber coupling}
\subsection{Coupling OAM into optical fibers}
To propagate OAM in an optical fiber, fiber eigenmodes with the same propagation constant should be excited in order to guarantee that no mode walk-off affects the OAM mode \cite{Yue}. RCFs with a single high refractive index ring can support the radially fundamental modes, making it an appropriate platform for OAM transmission. Although the use of RCF can maintain the OAM mode during propagation with reduced mode cross-talk, the high refractive index ring dimensions and doping properties will determine the maximum number of modes supported by the RCF. The OAM modes supported by  an optical fiber can be described as a linear combination of the linearly polarized (LP) modes \cite{Rojas:21, Zeng:18} or the vector modes \cite{Yue} supported by the fiber.

A set of OAM modes typically used in telecommunications are the LG modes characterised by index $p$ and $l$, these are radial index and OAM value, respectively \cite{Allen:92}. To couple OAM modes into optical fibers using LG modes, the radial number $p=0$ is used, which corresponds to an intensity profile consisting of a single ring \cite{Wang:12, Bozinovic:13}. Using these LG modes, however, leads to different coupling efficiencies for different $l$ values, since the ring radius of LG modes scales with $\sqrt{|l|} $\cite{Rojas:21}.  The PVB modes, on the other hand, have a ring-like intensity distribution whose radius varies much less as a function of the OAM values \cite{Vaity:15}. Therefore, using PVBs to couple OAM into fiber should have a higher overall efficiency for a given set of $l$ values compared to LG modes, as described in Ref. \cite{Rojas:21}.  In the next section we define these modes mathematically and describe techniques for their generation.      

\subsection{OAM generation using spatial light modulators}\label{sec:PVB-generation}

In order to generate a PVB, a real-time method capable of controlling every beam parameter is required. It was shown in Ref. \cite{Vaity:15} that it is possible to generate PVBs in free-space through a computer generated hologram (CGH) imprinted upon a spatial light modulator (SLM) and an optical lens in a Fourier transform configuration. Additionally, it is possible to implement a Fourier transformation using the CGH through Fresnel domain pattern of the hologram without needing a lens in the process\cite{Arrizon:07}. Thus, PVB generation can be totally controlled by a CGH. As in Ref. \cite{Arrizon:07}, creating a CGH for deploying on phase-only SLM is possible using the correct encoded of the complex scalar field, particularly the PVB. This was mathematically described in Ref. \cite{Vaity:15}:

\begin{equation}\label{eq:PVB}
    PVB_l(r,\theta)=i^{l-1}\left(\frac{\omega_g}{\omega_0}\right)e^{il\theta}e^{ \left( -\frac{r^2 +r_r ^2}{\omega_0 ^2}\right)} I_{l}\left( \frac{2r_{r}r}{\omega_{0}^{2}}\right),
\end{equation}
 where $r$ and $\theta$ are radial and azimuthal coordinates in a cylindrical system, $l$ the OAM topological charge, $\omega_0$ is the beam waist of the PVB beam, $\omega_g$ is the beam waist of the gaussian field incident on the SLM, $r_r$ is the ring radius of the PVB and $I_l(\cdot)$ is the $l$-th modified Bessel function from the first kind. 
 Additionally, $r_r$ can be defined in terms of the radial wavenumber ($k_r$) using $r_r=\frac{k_r f}{k}$, where $f$ is the focus of the lens that allows the Fourier transformation and $k$ is the wavenumber of the beam. The parameter $k_r$ can be used experimentally as an adjustable parameter to keep a constant ring radius from PVB mode\cite{Vaity:15}. 
\par
A similar technique can be used to generate LG beams. In this case, the transverse spatial profile of the complex scalar field is given by following equation in Ref. \cite{Allen:92}: 

\begin{multline}\label{eq:LG}
 \text{LG}_{p}^{l}(r,\theta,z)=\sqrt{\frac{2p!}{\pi (p+|l|)!}}\frac{1}{\omega(z)}\left(\frac{r\sqrt{2}}{\omega(z)}\right)^{|l|}\text{exp}\left(\frac{-r^2}{\omega(z)}\right)L_{p}^{|l|}\left(\frac{2r^2}{\omega^{2}(z)}\right)e^{il\theta}\\
 \text{exp}\left(-ik\frac{r^2 z}{2(z_{R}^{2}+z^{2})}\right)\text{exp}\left[-i(2p+|l|+1)\text{tan}^{-1}\left(\frac{z}{z_{R}}\right)\right],
\end{multline}
where $L_{p}^{|l|}$ is an associated Laguerre polynomial, with $l$ and $p$ the azimuthal and radial index, respectively, $z_R$ is the Rayleigh range and $\omega(z)$ is the beam waist (considered constant in this work). The $p$ parameter is associated with numbers of the concentric rings in transversal profile in LG modes. In this work we study OAM modes coupling into RCF with just one ring radius, then the radial parameter is fixed in $p=0$. Equation \eqref{eq:LG} is described using the same coordinates as the PVB (Eq.~\eqref{eq:PVB}) and adds $z$ coordinate in the beam propagation direction.

\section{Experimental setup}\label{Experimentalsetup}

\begin{figure}[ht]
        \centering
        \includegraphics[width=1\textwidth]{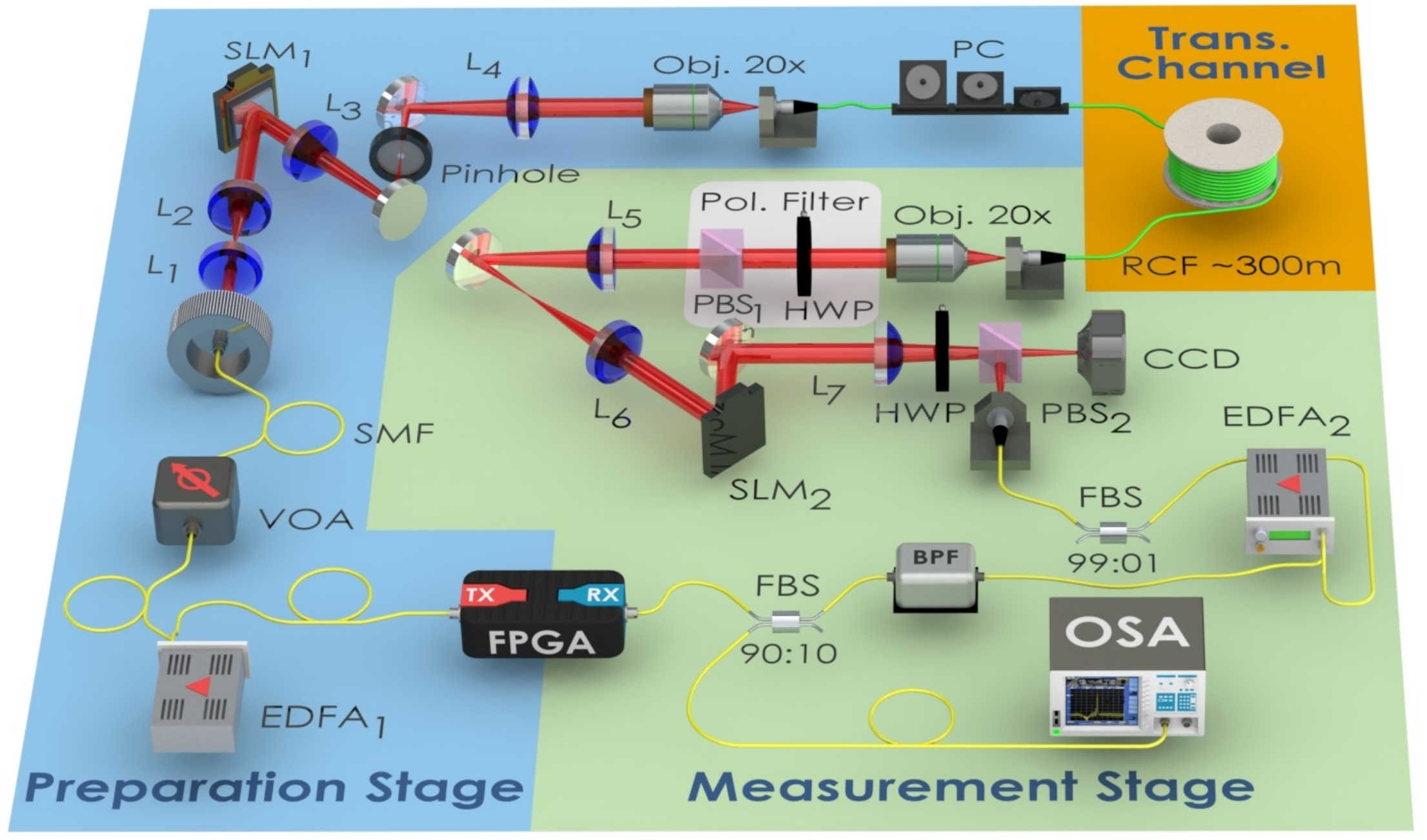}
        \caption{Experimental setup. Blue, orange and green backgrounds represent preparation, transmission and measurement stages, respectively.}
         \label{Fig_1}
\end{figure}
The experimental setup used in this work to generate, transmit and measure PVB is illustrated in Fig.~\ref{Fig_1}. Each stage is highlighted with a color, blue for preparation, orange for transmission and green for measurement. 
A commercial SFP+ transceiver was used to generate and detect an optical signal. The transceiver was driven by a field programmable gate array (FPGA) with  a pseudorandom binary sequence (PRBS) of length $2^{37}$. The transmitter uses an optical source centered at $1550\ nm$, which was amplified by an erbium-doped fiber amplifier ($\text{EDFA}_1$) and a variable optical attenuator (VOA) to control the signal power. Subsequently, the optical singal was launched into free-space, collimated and passed through to half wave plate (HWP) and polarizing beam splitter (PBS) to prepare the incident beam before the SLM, not shown in Fig.~\ref{Fig_1} for simplicity.
The OAM mode is generated when a horizontally polarized Gaussian beam  (with $\omega_g \approx1140\ \mu m$) is reflected by the phase-only SLM ($\text{SLM}_1$). $\text{SLM}_1$ deploys a CGH with the desired OAM mode. The beam reflected by $\text{SLM}_1$ passes through the $4f$ confocal optical system formed by two lenses ($\text{L}_3$ and $\text{L}_4$, with focus $f_3 =f_4 = 150\ mm$) and a pinhole placed between them. The pinhole filters the first diffraction order from the CGH, this way the generated OAM mode is placed in the image plane of this optical system.
The OAM mode is coupled into an RCF, used to propagate the signal. To couple the OAM mode a $20\text{x}$ objective lens is used. The RCF has an internal and external radius given by $a=6\ \mu m$ and $b=9\ \mu m$, respectively \cite{Long:18}. A total length of $300\ m$ of RFC was used. A RCF-based polarization controller was used to control the OAM mode coupled into the fiber.
After the OAM modes are propagated through the RCF, a second $20\text{x}$ objective lens was used to generate the image of the OAM-beam exiting the fiber core. This optical system magnifies  the mode radius $33 \times$ to achieve $r_r \approx230\ \mu m$. A polarization filter comprised of HWP followed by a PBS was used at the fiber output to ensure the horizontal polarization of the OAM-beam. A second $4f$ confocal optical system ($\text{L}_5$ and $\text{L}_6$) was used to magnify   the OAM-beam $2 \times$ after exiting of polarization filter, and it allows the image of OAM-beam to be produced on $\text{SLM}_2$. On $\text{SLM}_2$ we deployed a set of forked holograms (CGH of the LG basis), one at a time, encoding a conjugated OAM value relative to $\text{SLM}_1$ \cite{Pinnell:19}.
Finally, $\text{PBS}_2$ was used to select whether the signal was sent to a charge coupled device (CCD) camera or coupled into a SMF for detection. The SMF is placed at the center of the Fourier transform plane of the last lens ($\text{L}_7$). To detect the transmitted sequences, the signal was amplified by $\text{EDFA}_2$ and filtered to remove out-of-band amplified emission noise. An optical spectrum analyzer (OSA) was used to record the received spectrum and power. At the optical receiver, the FPGA was used to determine the bit error rate (BER). 

\section{Results}

Here we present the experimental results on the generation, fiber coupling and measurements of OAM. Additionally, we present results on two important application of OAM beams, fiber based communication systems and quantum system.

\subsection{Mode generation}

\begin{figure}[ht]
    \centering
    \includegraphics[width=\textwidth]{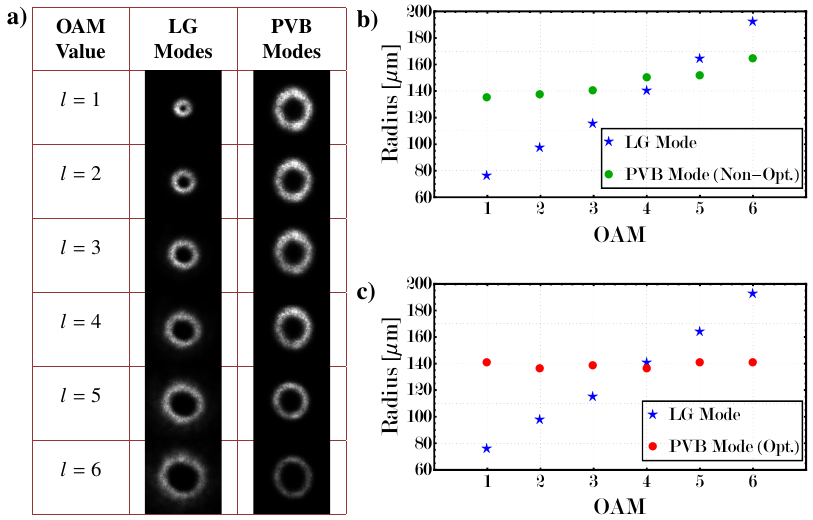}
    \caption{a) Transversal intensity profiles of the generated OAM modes for LG (second column) and PVB modes (third column). b) and c) represent two graphical comparisons of PVB modes ring radius values with ring radius values from LG modes (blue-star-points in both cases) as function of the $l$ OAM values. Where b) compares LG modes with PVB non-optimized modes ($k_r$ fixed, green-dot-points), and c) compares LG modes with PVB optimized modes ($k_r$ variable, red-dot-points).}
    \label{Fig:combined}
\end{figure}

To study the OAM mode generation we generated PVB and conventional LG beams using the SLM together with the $4f$ optical system. Using a CCD camera the intensity of the generated beams was recorded in the image plane of the first $4f$ optical system ($\text{L}_3$ and $\text{L}_4$) before the fiber coupling system (see Fig.~\ref{Fig_1}). Results of the OAM generated using a CGH on the SLM are shown in Fig. \ref{Fig:combined}a, where the detected intensity for each mode order $l$ is shown. 
The ring radius of the PVB are shown to remain constant as the order $l$ is increased from $l=1$ to $l=6$, while the LG beam presents a large radius variation. The measured ring radius for the PVB and LG modes is shown Fig.~\ref{Fig:combined}b, where the LG modes presented ring radius variation from $77\ \mu m$ to $193\ \mu m$. 
An average ring radius of $147.9\ \mu m$ was observed when the $k_r$ parameter was set to $k_r =3.06416$. A variation of $29\ \mu m$ in the ring radius was observed in this scenario, with the largest radius of $165.9\ \mu m$ observed for $l=6$. 
Next, we studied the impact of the $k_r$ parameter on the ring radius of the generated PVB, with the results shown in Fig.~\ref{Fig:combined}c. 
The optimization was performed based on $k_r =3.06416$ obtained for $l=3$. Then, $k_r$ was either increased or decreased to maintain a fixed ring radius using Eq.~\eqref{eq:PVB}. In this scenario, an average ring radius of $140.5\ \mu m$ was obtained with a maximum variation of $4.7\ \mu m$.
Note that the differences reported with and without the $k_r$ optimization are significantly less compared to the ring radius variation observed for the LG modes. 
For example, the ring radius variance of LG modes was approximately $1873.3\ \mu m^2$, while the ring radius variances measured for PVB modes were of $5.4\ \mu m^2$ and $122.1\ \mu m^2$ with $k_r$ and without $k_r$ optimization, respectively. Therefore, the LG modes have a variance $15$ times higher than the PVB modes without adjusting $k_r$. In the case of LG modes, it is well known that their radius grows with $\sqrt{|l|}$, and in order to keep their ring radius constant a different optical system should be used for each OAM.  

\subsection{Fiber coupling and propagation}

\begin{figure}[ht]
        \centering
        \includegraphics[scale=1]{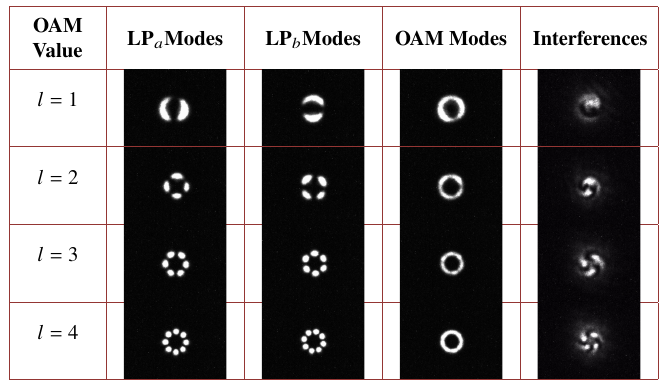}
        \caption{Modes propagated through RCF. First column shows $LP_a$ modes, second column shows $LP_b$ modes, third column shows OAM modes, and fourth column shows interference pattern for $l=1$ to $l=4$.} 
         \label{fig:OAM_measureredquality}
\end{figure}

\begin{figure}[ht]
        \centering
        \includegraphics[scale=1]{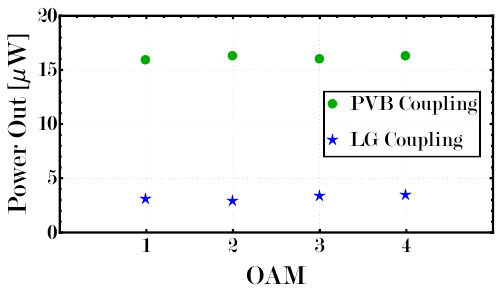}
        \caption{Output power measurements from RCF at the image plane of the second objective lens $20\text{x}$ using different coupling modes, PVB modes (green-dot-points) and LG modes (blue-star-points), as a function of OAM value.} 
         \label{Fig: OAM output power}
\end{figure}

OAM modes in an optical fiber can be described using the linear combination of LP modes, as in \cite{Rojas:21, Yue, Zeng:18}. We have used a GRIN-RCF with inner and outer ring radius of $6\ \mu m$ and $9\ \mu m$ \cite{Long:18}. The RCF supports propagation of OAM modes up to $l=4$. Considering the fixed ring radius of the RCF, we used $r_r \approx140\ \mu m$ for the generated PVB modes in order to maintain a constant $k_r$ parameter for all modes.
At the fiber input a polarization controller (PC) was used to control the linear combinations of LP modes in the fiber. At the fiber output the following parameters were evaluated: field intensity, interference between OAM and a Gaussian mode, and optical power. The results are summarized in Fig.~\ref{fig:OAM_measureredquality} and Fig.~\ref{Fig: OAM output power}. Fig.~\ref{fig:OAM_measureredquality} shows the intensity profiles at the fiber output for the $LP_{a,b}$ modes excited by the OAM modes at the fiber input. This was achieved using the polarization controller to excite $LP_a$, $LP_b$, or an OAM mode from the bend perturbation to the field transmitted through the RCF \cite{Dashti:06, Gregg:15, Ma:21}. Note that the fiber modes propagated are independent of the free-space beam used to excite them, the difference arises from the coupled mode intensity. Additionally, the transmitted OAM  out of the RCF is shown together with the OAM-Gaussian beam interference pattern that shows successful OAM propagation \cite{Jiang:17, Willner:15, Vaity:15, Heckenberg:92, Andrews:15}.    

The coupling efficiency between the OAM modes generated with the CGH and the RCF was characterized using the received power at the fiber output. The optical power was measured at the image plane of the second objective lens ($20\text{x}$, see Fig. \ref{Fig_1}) using a constant coupling power into the RCF for both OAM modes (LG and PVB, $\text{P}_{\text{in}}\approx450\ \mu W$). An average output power of $\text{P}_{\text{out}}^{\text{PVB}}\approx16\ \mu W$ was measured for the PVB modes, while an average output power of $\text{P}_{\text{out}}^{\text{LG}}\approx3.32\ \mu W$ was observed for LG modes. These results show the higher coupling efficiency of the PVB modes compared to the LG modes. 
To verify the OAM modes propagated through the RCF we used an interferometric technique between the OAM mode and a Gaussian beam \cite{Jiang:17, Vaity:15, Willner:15}. The interference patterns shown in Fig.\ref{fig:OAM_measureredquality}, verify qualitatively that the OAM modes have effectively propagated through the RCF and have $l$ values of $l={1,2,3,4}$.   

\subsubsection{Classical communication systems}

\begin{figure}[ht]
    \centering
    \includegraphics[scale=1]{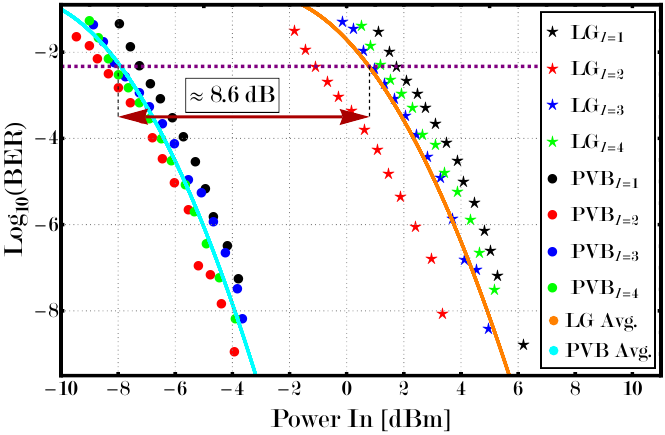}
    \caption{Measured BER after transmission in the RCF using PVB (dot-points) and LG (star-points) modes. The continuous lines, represent average of $log(BER)$ values using PVB (cyan) and LG (orange) modes. The dashed purple horizontal line represents HD-FEC threshold.}
    \label{Fig:BER}
\end{figure}
As described above, OAM was proposed as a way to implement SDM to increase the capacity of optical fiber communication systems. Considering this, the use of PVB to couple OAM into optical fibers can have a positive impact on the systems, allowing to transmit longer distances compared to LG modes and simplify the design for a OAM coupling system.
Here, the quality of the transmitted signals was evaluated using a commercial  SFP+ transponder operating at $1$~Gbps. The bit error rate (BER) was measured after fiber propagation. Fig.~\ref{Fig:BER} shows the measured BER as a function of the power coupled into the RCF for the PVB and LG mode using $l={1,2,3,4}$.
PVB are shown to require lower coupling power to achieve a fixed BER of $4.7x10^{-3}$ (FEC limit using a a $7\%$ FEC overhead \cite{Zhang:14}) compared to the LG modes. A power difference of $8.6$~dB was observed for the mean BER between PVB and LG modes. Additionally, the PVB modes are shown to have smaller performance difference for different $l$ values at a fixed BER, with a $0.6$~dB difference observed, compared to the $1.2$~dB measured for the LG modes. 

From the Fig.~\ref{Fig:BER}, we can see that the best performance was obtained using PVB with $l=2$, because its needs the lowest optical power to transmit the $1$~Gbps signal. This characterization shows that the PVBs are better candidates to couple amplitude modulated information in fiber-based SDM systems, allowing to implement systems with similar power requirements.

\subsubsection{Quantum applications}
In the quantum regime, it is well known that some quantum information protocols (QIPs) can be more robust when using quantum states in higher dimensions, these are quantum systems with d$>2$ levels (qudit) \cite{Gisin:02,walborn08b, Pironio:10, Gallego:10, Daniel:18, Aguilar:18}. Photons carrying OAM can be used to encode OAM-qudits, thus allowing more information to be transmitted between the communicating parties \cite{Yao:11, Cerf:02, Mirhosseini:15}. OAM-qudits have a myriad of applications in QIPs, both in free space and optical fiber channels, most notably quantum key distribution (QKD) protocols \cite{Nicolas:15, Erhard:17, Sit:18, Cozzolino:19, Otte:20}. In particular, RCF has succeeded in this context with demonstrations showing that the OAM modes supported by RCF can be used to enhance classical and quantum communications links \cite{Bozinovic:13, Nejad:16, Long:18, Cozzolino:19, Sit:18, Cao:20, Xavier:20, Caas:21}.

As shown in Ref.~\cite{Rojas:21}, asymmetric coupling of OAM beams into RCF leads to undesired state transformations performed by the RCF, which results in high quantum bit error rates (QBER). To overcome this drawback PVBs are proposed \cite{Rojas:21}. The QBER is related to the average fidelity of the quantum states used for a QKD protocol, as follows $\Bar{F}=1-$ QBER. Here we use PVBs to experimentally demonstrate that  4-dimensional OAM-encoded quantum systems can be propagated through the $300$~m of RCF with high fidelity. As reported in Ref.~\cite{Cerf:02}, to guarantee the unconditional security of QKD protocols, we must be below the upper limit of the QBER for the coherent attacks ($D^{coh}$), which depends on the dimension of the quantum system used. In this case, for 4-dimensional OAM-encoded quantum systems, the upper limit of the coherent attack is $D^{coh}=18.96\%$. Thus the average fidelity must be greater than $\Bar{F}=0.81$.

\begin{figure}[ht]
        \centering
        \includegraphics[width=\textwidth]{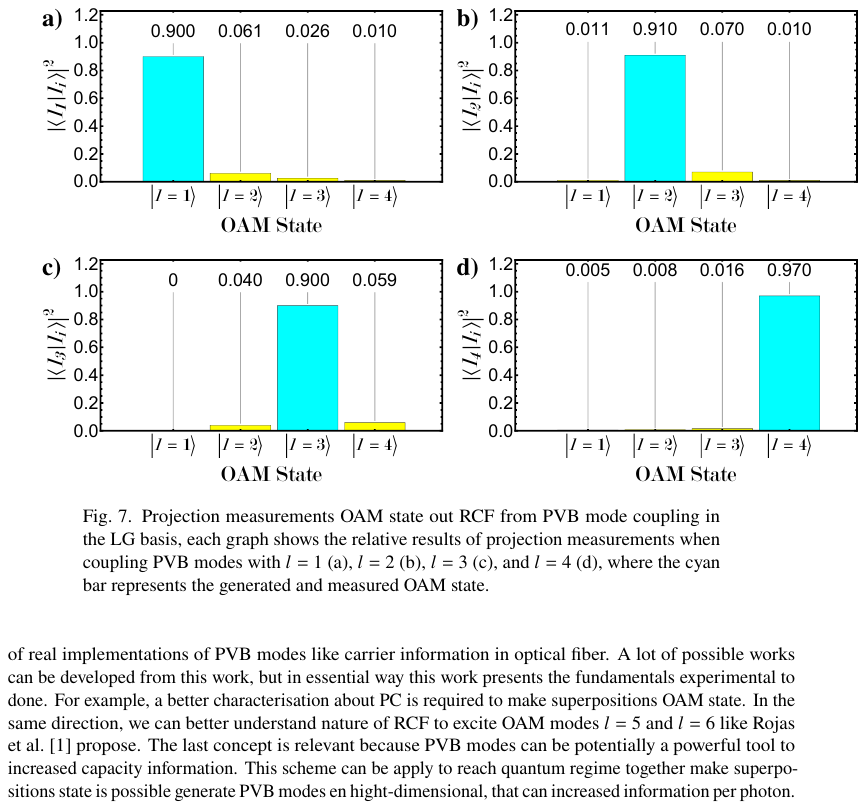}
        \caption{Four measured fidelities after propagated through the $300$~m of RCF, for the states a) $|l=1\rangle$, b) $|l=2\rangle$, c) $|l=3\rangle$, and d) $|l=4\rangle$. The average fidelity the states is $92\%$.} 
         \label{fig:quantum_fidelity}
\end{figure}

Fig.~\ref{fig:quantum_fidelity} shows the measured fidelity of the 4-dimensional OAM-encoded quantum states after the propagation through the RCF with the average fidelity of $92\%$. This paves the way for implementing QKD protocols using high-dimensional OAM-encoded quantum states in SDM optical fibers.

\section{Conclusion}
Here we experimentally studied the generation and transmission properties of two OAM carrying beams in ring core optical fibers. The beams studied were the perfect vortex beams (PVB) and Laguerre-Gaussian (LG) beams. 
We have shown that when using a single preparation stage for all of the OAM beams, the PVBs present less ring radius variation and are able to couple higher powers into a ring core fiber. Ring radius variations from $77\ \mu m$ to $193\ \mu m$ where observed for LG beams, while an average ring radius of $140.5\ \mu m$ with a maximum variation of $4.7 \mu m$ was observed for the PVBs.
Additionally, higher optical powers were coupled into the RCF using PVBs. An average coupled optical power of $15\ \mu W$ and $4\ \mu W$ were observed for PVB and LG modes, respectively.
In addition to successfully coupling optical OAM into a RCF using PVBs, we have also demonstrated the possibility of establishing a transmission link operating at $1$~Gbps. The results show that the high coupling powers of PVB lead to lower power requirements to establish fiber-based communications links. An average $8.6$~dB difference in required input power was observed using PVBs compared to LG modes.
In addition, we have shown that the constant coupling power of PVBs as a function of the mode order $l$ can allow for the implementation of QKD protocols. Here the figure of merit was the measured quantum state fidelity. We report an average fidelity of $92 \%$ in a 4-dimensional quantum system after transmission in the RCF, which is larger than the minimum fidelity required for successful key distribution under coherent attacks.
Our work therefore can be seen as an initial step for the development of PVBs as information carriers in optical fibers leading SDM systems in telecommunications, and to implement high dimensional QKD systems. In both scenarios, more efficient coupling of OAM into ring-core fibers is a powerful tool to increased information capacity.


\section*{Funding} The authors acknowledge support from Fondo Nacional de Desarrollo Científico y Tecnológico (ANID) (Grants No. 1190933, 1200266, 1200859, 1220960, 1231826) and ANID – Millennium Science Initiative Program – ICN17\_012. Also, this work was supported by Vetenskapsrådet (VR) under the project “Scaling Quantum Networks by Multi-dimensional Multiplexing,” and the National Natural Science Foundation of China (62225110).


\section*{Acknowledgments}
The authors thank G. B. Xavier for valuable discussions.

\section*{Disclosures} 
The authors declare no conflicts of interest.

\section*{Data Availability Statement}
Data underlying the results presented in this paper are not publicly available at this time but may be obtained from the authors upon reasonable request

\bibliographystyle{unsrt}
\bibliography{sample}
\end{document}